# A method for reduction of propagation loss of surface plasmons. Experimental demonstration of the loss reduction for Fe/MgO/AlGaAs plasmonic structure integrated with AlGaAs/GaAs optical waveguide

V. Zayets, H. Saito, K. Ando, and S. Yuasa

Spintronics Research Center,
National Institute of Advanced Industrial Science and Technology (AIST), Japan

*Abstract:*

*A method for the substantial reduction of propagation loss of surface plasmons was proposed and experimentally demonstrated. The method is based on the fact that the propagation loss of the surface plasmons depends significantly on the optical confinement of the plasmon. A plasmonic structure, which contains a metal and two dielectric layers of different refractive indexes, is proposed in order to optimize optical confinement and to reduce propagation loss of the surface plasmons. A low propagation loss of 0.17 dB/µm for a surface plasmon in a Fe/MgO/AlGaAs plasmonic structure was achieved. A good coupling efficiency of 2.2 dB/facet between a surface plasmon in Fe/MgO/AlGaAs and a waveguide mode in AlGaAs/GaAs optical waveguide was demonstrated.*

A new design of an integrated optical isolator, which utilize unique non-reciprocal properties of surface plasmons, has been proposed[1]. Main obstacle for a practical realization of the proposed design is a substantial propagation loss of the surface plasmons in structures containing a ferromagnetic metal. The reduction of the propagation loss of a surface plasmon is a key to make the plasmonic isolator competitive with other designs of the integrated isolator.

The optical isolator is an essential component of optical communication systems. It protects optical components from unwanted back reflections. In optical fiber networks a bulk-type optical isolator made of a magnetic garnet is used. The optical bulk-type isolator with the high isolation ratio of 40 dB, the low insertion loss of 0.7 dB and the wide bandwidth of 20 nm is commercially available. The integration of optical elements into a Photonic Integrated Circuit (PIC) is an important task, because it may reduce a cost and improve performance of high-speed optical data processing circuits. The integration of an optical isolator is important for PICs, because the problem of back-reflected light is more severe in the case of integrated optical elements. However, the optical isolator is one of the few optical components, which has not yet been integrated into commercial PICs.

The reason, why it is difficult to integrate the optical isolator into a PIC is the following. The indispensable component of any optical isolator is a magneto-optical (MO) material. Traditionally, the magnetic garnets are used as a transparent MO material for the optical isolator[2-10]. However, it is difficult to grow the required high-crystal-quality magnetic garnets on semiconductors substrates Si, GaAs, InP, which are used as a substrate for a PIC. It is because the conventional growth temperature of the garnets is above the melting temperature of these substrates. The recent significant improvement of the growth technique for the growth of the magnetic garnets on semiconductor substrates should be noted[11-13].

To overcome this problem several solutions have been proposed. A wafer bonding between garnets and semiconductors[8-10] and sputtering of garnet on Si[11-13] were successfully used to fabricate the isolator on a Si substrate. Diluted magnetic semiconductor CdMnTe[14] have been used for the fabrication of an optical isolator on a GaAs substrate. An optical amplifier with a ferromagnetic electrode operating as an optical isolator was demonstrated[15-20]. Even though a good performance of the above-mentioned standalone isolators fabricated on the semiconductor substrates was achieved, the integration of the isolators with other optical elements into the PIC has not been demonstrated yet. Major difficulties of the integration were the technological compatibility of the isolator with other optical components of the PIC or a high insertion loss or a long isolator length or a narrow operational bandwidth or a high operational current.

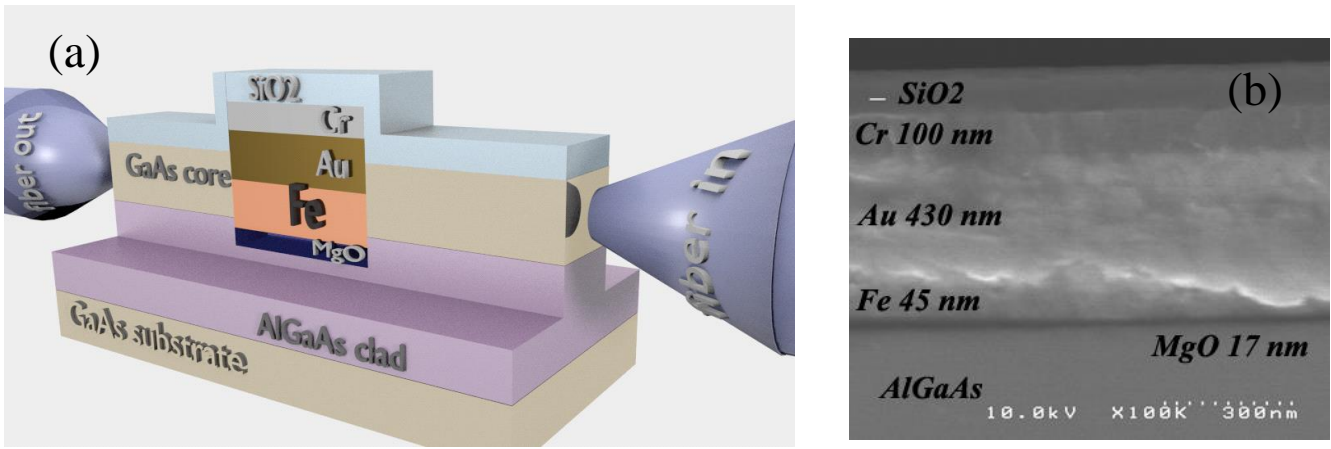

*Fig. 1. (a) Plasmonic isolator. It consists of a stack of metallic layers embedded into AlGaAs/GaAs optical waveguide. The magnetization of Fe is in-plane and it is perpendicular to the light propagation direction; (b) Cross-sectional SEM image of the plasmonic section.*

In a plasmonic isolator, a ferromagnetic metal is used as a MO material. The advantages of the plasmonic isolator is a good technological compatibility with the fabrication technology of the PIC and its short length of a few micrometers. The plasmonic isolator can be fabricated using conventional sputtering and lift-off techniques. These techniques are already used for the fabrication of other components of the PIC and the additional technological steps for the fabrication of the plasmonic isolator should not disturb the existing technology of the PIC fabrication. The length of the plasmonic isolator is only about 4-8 µm. There is no other isolator design with a comparably short isolator length. For example, the length of the above-mentioned integrated isolators varies from a millimeter to a few centimeters. The plasmonic isolator may be a good option as an optical isolator for dense-integrated circuits.

Figure 1 (a) shows the design of the plasmonic isolator, which is studied in this Letter. It consists of a stack of metallic layers embedded into an AlGaAs/GaAs optical waveguide. The metallic stack fully blocks the direct light propagation from the input waveguide to the output waveguide. Light can reach the output only when a surface plasmon is exited at the Fe/MgO surface. Because of the transverse non-reciprocal magneto-optical effect[21] in the iron, the propagation loss of the surface plasmons may be significantly different for the forward and backward propagation. In the case when in one direction the loss is small and in the opposite direction the loss is large, this simple structure operates as an optical isolator.

The non-reciprocal function of the plasmonic isolator can be achieved only in a magneto-optical material and ferromagnetic metals like Fe, Co or Ni have to be used in the plasmonic isolator. It is known[1,21-23] that the propagation loss of the surface plasmons in these ferromagnetic metals is at least an order larger than the optical loss of plasmons in Au, Ag and Cu, which are the conventional metals for the plasmonic devices. The high propagation loss is the reason why the long-range surface plasmons have not been observed directly on a surface of a ferromagnetic metal and a thick buffer layer of a noble metal had to be used between a ferromagnetic metal and a dielectric[22,23] in order to observe the long-range surface plasmons.

Ideally, an optical isolator should not have any optical loss in the forward direction. For the PIC the forward loss of a few dB can be acceptable. An optical isolator with a higher optical loss can not be used in the PIC. Therefore, the reduction of the insertion loss is the highest-priority task for the developing of the plasmonic isolator.

In this Letter we propose a method for reduction of propagation loss of a surface plasmon. We propose a new design of plasmonic structure, which contains a metal and two dielectric layers of different refractive indexes. By optimizing this structure, it is possible to reduce significantly the optical loss of the surface plasmons. Using this method, we experimentally demonstrate a long-distance propagation of a surface plasmon on surface of a ferromagnetic metal. The low propagation loss of 0.17 dB/µm was achieved in a Fe/MgO/AlGAAs plasmonic structure. The propagation loss is reduced 10-15 times compared to a conventional single-dielectric-layer Fe/AlGAAs plasmonic structure.

A unique property of surface plasmons is that the optical confinement and the propagation loss of a surface plasmon can be controlled and modulated by a slight change of the conditions at the metal-dielectric interface. Figure 2 explains this fact by comparing the field distributions of surface plasmons and the field distribution of a mode of an optical waveguide. In the optical waveguide (Fig.2(a)), light propagates inside the core layer reflecting from the clad and cover layers of a lower refractive indexes. Light is confined inside the core layer. This means the approximate size of the light beam is fixed by the thickness of the core layer. Even if the refractive index of the waveguide layers may change, the approximate size of the light beam does not change and it is slighter wider than the thickness of the core layer. This means that a waveguide mode is confined by the volume of the core layer. In contrast, a surface plasmon is confined by a surface between a metal and a dielectric. This means the size of the surface plasmon is not fixed by any geometrical sizes, but it only depends on the structure of the interface between the metal and dielectric.

Since a metal absorbs light, the less light is inside the metal and the more light is inside the dielectric, the smaller the loss of the surface plasmons is. The penetration depth of light into a metal is called the skin depth. The skin depth is a parameter of a metal and it practically does not depend on the parameters of the plasmonic waveguide such as the refractive index of the dielectric or the structure of the layers of the dielectric. In contrast, the penetration depth of light into the dielectric depends significantly on those parameters. It may vary from a few tens of micrometers to a few tens of nanometers. In the case of a looser confinement of a surface plasmon (Fig.2 (c)), most of light propagates inside the dielectric and the surface plasmon experiences a smaller propagation loss. In the case of a tighter confinement of a surface plasmon (Fig.2 (b)), there is a smaller amount of light in the dielectric and the propagation loss of the surface plasmon becomes larger.

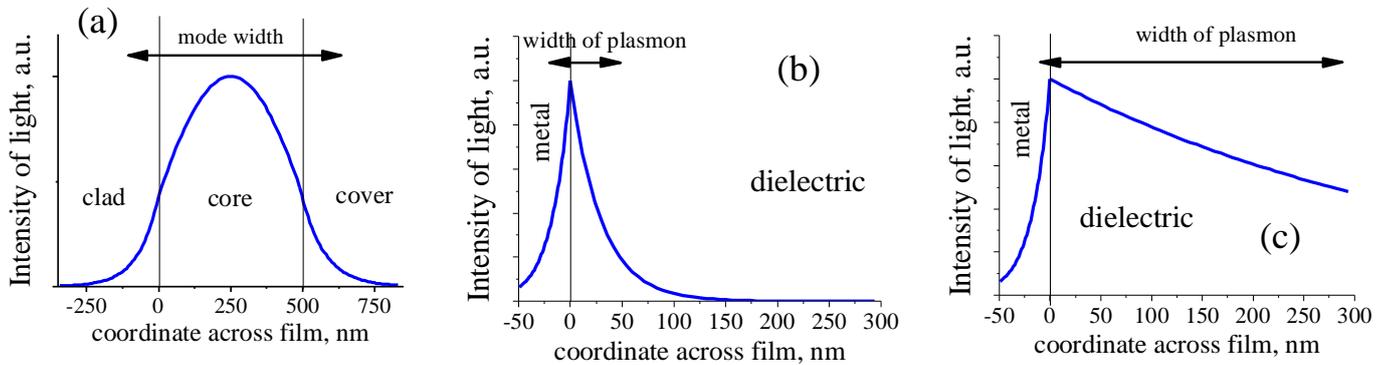

Fig.2 Cross-sectional distribution of the optical field of (a) a mode of an optical waveguide (b) a surface plasmon with a tight confinement; (c) a surface plasmon with a loose confinement. The arrows show the effective width of the optical mode and the surface plasmon. The effective width is determined as width between points of the 1/e decrease if the intensity. The effective width of a waveguide mode is slightly wider than the width of core layer. The width of a plasmon is not fixed by any geometrical size.

Two different types of applications of the surface plasmons can be distinguished. There are applications where focusing of a laser beam into a smallest-possible spots is required. In this case the surface plasmon with a tight confinement is a good option. Using this type of the surface plasmons, it is possible to achieve a spot size as small as ~10-20 nanometers[24]. Another possible application is the use of surface plasmons in data-processing devices. In these devices a low insertion optical loss is a critical parameter and the surface plasmons with loose confinement should be used.

There are several methods to control the confinement of the surface plasmons. One method is to insert at the metal/dielectric interface a thin layer of another dielectric with a different refractive index[1,21]. For example, a MgO layer (n=1.71 at λ=1550 nm) can be inserted between $Al_{0.5}Ga_{0.5}As$ (n=3.04 at λ=1550 nm) and a metal. Figure 3 (a) shows the calculated penetration depth into the dielectric and the loss of the surface plasmons in the $Al_{0.5}Ga_{0.5}As$/MgO/Fe plasmonic structure as a function of the MgO thickness. The calculations were done using the rigorous solution of the Maxwell equations[1,21]. As the MgO layer becomes thicker, the 1/e penetration depth increases and the propagation loss of the surface plasmons decreases. This means that the insertion of the MgO layer reduces the optical confinement for a surface plasmon. A thicker MgO layer makes the conferment weaker. For the MgO thickness near 17 nm, the propagation loss approaches to zero and the 1/e penetration depth sharply increases. This means that the size of the surface plasmon becomes larger and the plasmon confinement becomes weaker. The structure with the MgO thickness greater than 17 nm does not support surface plasmons. The MgO thickness of 17 nm is defined as the cutoff thickness. Structures with the MgO thickness thinner than cutoff thickness provide sufficient confinement for the propagation of a surface plasmon. When the MgO layer is thicker, the confinement is not sufficient and a surface plasmon can not propagate.

In the case of a thick MgO layer, the influence of the $Al_{0.5}Ga_{0.5}As$ layer becomes weak and the structure becomes similar to a conventional single-dielectric plasmonic structure MgO/Fe, which supports a surface plasmon. Therefore, the $Al_{0.5}Ga_{0.5}As$/MgO/Fe structure with a sufficiently thick MgO layer should also support a surface plasmon. Figure 3(b) shows the calculated 1/e plasmon propagation distance and the magneto-optical (MO) Figure-of-merit as a function of the MgO thickness. The MO Figure-of-merit (FoM) is the ration of the non-reciprocal loss to the average loss[1,15,21]. As can be seen from Fig. 3(b), there is another cutoff thickness of 480 nm. For the MgO layer thicker than this cutoff thickness, the confinement is sufficient for a surface plasmon to propagate.

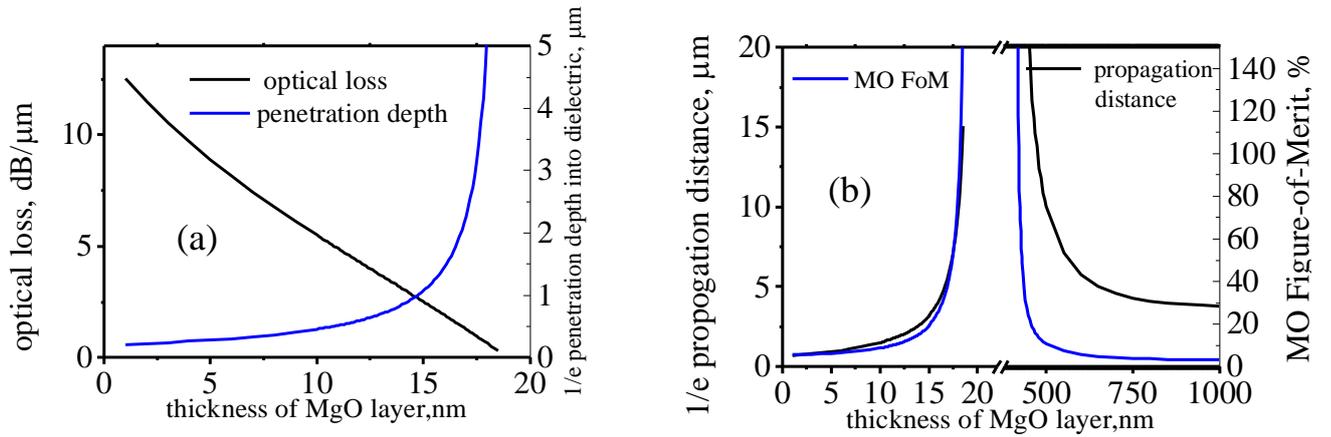

Fig.3. Surface plasmon in $Al_{0.5}Ga_{0.5}As$/MgO/Fe. (a) Propagation loss and 1/e penetration length into the dielectric. (b) 1/e propagation distance and MO Figure-of-Merit. λ=1550 nm

Near the cutoff the surface plasmons have several merits. They have a small propagation loss and a long propagation distance. Additionally, the surface plasmons can be efficiently modulated or be switched by a small variation of the refractive index of the metal or the dielectric. For example, a small change of the refractive index of the metal due to the magneto-optical effect may significantly change the propagation loss of surface plasmons. In the region near the cutoff the MO Figure-of-Merit can be enhanced[1,21] to 100-150% compared to 2-4% in regions far from the cutoff (Fig. 3(b))

In this Letter we demonstrate experimentally that the proposed method of the double-dielectric plasmonic structure is efficient for the reduction of the propagation loss of the surface plasmons. We integrated monolithically an AlGaAs/GaAs passive waveguide with a Fe/MgO/AlGaAs plasmonic structure on a GaAs substrate. The AlGaAs/GaAs waveguide was grown by the Molecular Beam Epitaxy (MBE) on a GaAs (100) substrate. Following a-hydrogen cleaning of the GaAs substrate, 3 μm of the $Al_{0.5}Ga_{0.5}As$ cladding layer and 450 nm of the GaAs cladding layer were grown. Next, a 3 μm- wide rib waveguides were fabricated by the wet etching out the GaAs top layer aside of the waveguide rib. The pattern for the plasmon structure was precisely aligned to the waveguide. In the pattern there was a hole in the intended place of the plasmonic section. The GaAs top layer was wet etched through the hole. The stack of MgO( 17 nm)/Fe (45 nm)/Au(430 nm)/Cr(100 nm) was deposited and was lifted-off. Next, 100 nm of SiO2 was deposited to avoid a metal oxidation. A thick Au layer was used in order to block any direct propagation of light from the input to the output waveguides. The Cr layer was used to prevent the propagation of the plasmons on top of the Au layer.

The sample was cleaved into 1-mm long waveguides and the fiber-to-fiber transmission was measured (Fig.4). The TM0 waveguide mode was excited from the input lensed polarization-maintaining fiber. There are several contributions to the fiber-to-fiber optical loss. The fiber-to-waveguide coupling loss is measured to be about 8.5 dB per a facet. The coupling loss between a waveguide mode and a surface plasmon is measured to be 2.2 dB per a facet. The propagation loss in the input and output optical waveguides was measured to be smaller than 0.1 dB. Additionally, there is a propagation loss of the surface plasmon. In order to measure the propagation loss of surface plasmons, the samples with different length of the plasmonic section were fabricated. Figure 4 shows the measured the fiber-to-fiber transmission for propagation length of the surface plasmons of 4 μm and 64 μm. The difference in transmission was about 10 dB, which indicates that the propagation loss of surface plasmon is 0.17 dB/μm. The same value of the propagation loss was measured for plasmon-propagation lengths of 8, 16, and 32 μm.

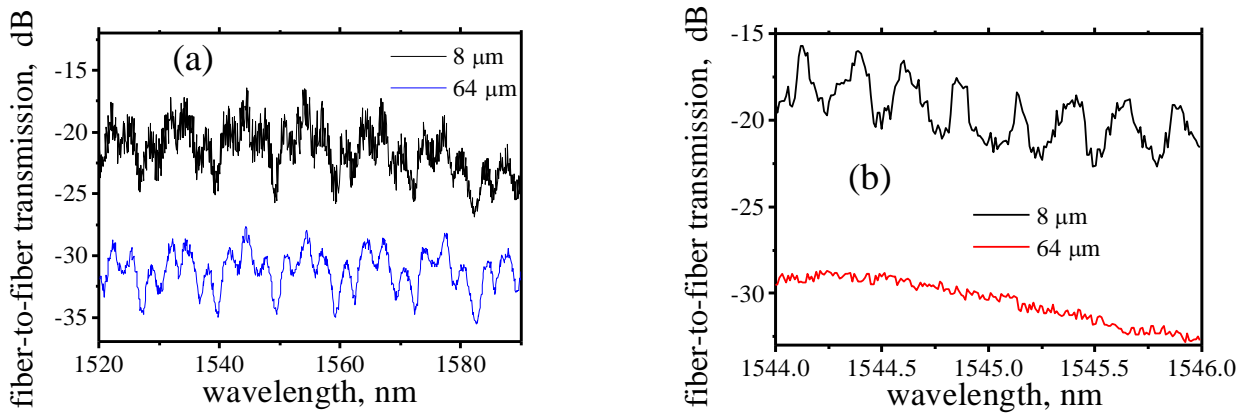

*Fig.4. Fiber-to-fiber transmission of the AlGaAs/MgO/Fe plasmonic structure embedded into the AlGAs/GaAs optical waveguide. TM0 waveguide mode was excited.*

The oscillations with the period of 3.3 nm and the amplitude of 2 dB are due to the multi reflections of a waveguide mode between waveguide interfaces with the fiber and with the plasmonic section. By fitting these oscillations, the coupling loss of 2.2 dB/facet between a waveguide mode and a surface plasmon is calculated.

A good coupling between a waveguide mode and a surface plasmon is a required parameter for any plasmonic device to be used in the PIC. All devices in the PIC are connected by optical waveguides and a bad coupling would cause an unacceptable substantial insertion loss of a plasmonic device. The optical field distribution is very different between a waveguide mode and a surface plasmon and matching them is a challenging task. The following method was used to match the field distribution of a waveguide mode and a surface plasmon. We have designed the optical waveguide such that the propagation conditions of the TM0 waveguide mode are near the cutoff conditions. If the cutoff thickness of the core layer for the TM0 mode was calculated to be 400 nm, we used waveguides with the core thickness of 450 nm. In this case there is a significant amount of light inside the AlGaAs cladding layer and the field distribution of the waveguide mode becomes similar to the field distribution of the surface plasmon. The obtained low coupling loss of 2.2 dB/facet demonstrates the effectiveness of this matching method. For example, the coupling loss between a surface plasmon and a waveguide mode of a conventional waveguide with thicker core layer exceeds 8 dB/facet.

Figure 4(b) shows the fiber-to-fiber transmission in the wavelength range from 1544 nm to 1546 nm. It is the same data as it is shown in Fig. 4 (a), but it is zoomed in near λ=1545 nm. In the case of the 8-μm-long plasmon structure, there are Fabry-Perot oscillations with the period of 0.25 nm. These Fabry-Perot oscillations occur because of the light reflections between the input and output waveguide interfaces. The existence of these Fabry-Perot oscillations is another proof of the low propagation loss of a surface plasmon and the good coupling efficiency between a surface plasmon and a waveguide mode. On the circle path between these interfaces, light experiences twice the propagation loss of the surface plasmons, twice the coupling loss between a waveguide mode and a surface plasmon and twice the loss when reflected at a waveguide/fiber interface. Still after this circle path, the intensity of the light is sufficient to interfere with input light in order to build the Fabry-Perot oscillations.

The measured MO Figure-of-Merit (FoM) was about 1-2 %, which is significantly smaller than expected (Fig. 2(a)). The reason for the small MO Figure-of-Merit is understood as follows. As can be seen from Fig. (2(a)), in order to obtain a large MO Figure-of-Merit the thickness of the MgO layer should be kept with a high tolerance. The thickness tolerance should be at least 0.1 %. This is not the case for our fabricated sample. Since the plasmonic structure was fabricated after the fabrication of the optical waveguides and the plasmonic structure was fabricated using the lift-off technique, the required smoothness of the MgO layer has not been achieved. As was measured from a cross-sectional SEM image (Fig. 1(b)), there is about 7 % variations in the MgO thickness between the center and the edges of the plasmonic structure and there are about 2 % short-distance variations of the MgO thickness. Another indication of insufficient tolerance of the MgO thickness is that in the fiber-to-fiber transmission spectrum there is no sharp resonance-like features as it is predicted (Fig.4(a)).

Even though the perfection of the interface was not sufficient to achieve a high MO Figure-of-Merit, the insertion of the MgO layer was efficient for the reduction of the propagation loss of a surface plasmon. As was explained above, the propagation loss of a surface plasmon is directly related to the confinement of the plasmon. As the MgO layer makes the confinement of the surface plasmon looser, the propagation loss decreases. That is why the tolerance of the MgO layer is not as critical for the loss reduction.

In conclusion, we have proposed a method for the reduction of the propagation loss of the surface plasmons. The method is based on the fact that the propagation loss of the surface plasmons depends significantly on the confinement of the plasmon. A surface plasmon with looser confinement has a smaller propagation loss. A unique property of a surface plasmon is that its confinement and therefore its propagation loss can be controlled and/or be modulated in a wide range by changing the refractive index of the dielectric or a layer structure of the dielectric near a metal-dielectric interface. Near the cutoff conditions, even a slight change of the refractive index of the metal or the dielectric may cause a significant change of the plasmon propagation loss. Inserting a thin dielectric layer at the metal-dielectric interface is proved to be an efficient method for the reduction of the propagation loss. Using the proposed method we have demonstrated experimentally a long-distance propagation in the AlGaAs/MgO/Fe plasmonic isolator. Simultaneously, we have demonstrated a good 2.2 dB/facet coupling efficiency between a waveguide mode and a surface plasmon. The proposed method of propagation-loss reduction is tolerant to imperfections of the fabrication technology. The perfection of the fabrication technology is crucial to achieve an expected large MO Figure-of-Merit for the surface plasmons. The proposed method for reduction of the plasmon propagation loss must be efficient not only in a MO plasmonic structure, but in other plasmonic devices where the lowest plasmon propagation loss is required.